# Impermeable Atomic Membranes from Graphene Sheets

*J. Scott Bunch, Scott S. Verbridge, Jonathan S. Alden, Arend M. van der Zande, Jeevak M. Parpia,*

*Harold G. Craighead, and Paul L. McEuen\**

Cornell Center for Materials Research, Cornell University, Ithaca, N.Y. 14853

mceuen@ccmr.cornell.edu

**We demonstrate that a monolayer graphene membrane is impermeable to standard gases including helium. By applying a pressure difference across the membrane, we measure both the elastic constants and the mass of a single layer of graphene. This pressurized graphene membrane is the world's thinnest balloon and provides a unique separation barrier between 2 distinct regions that is only one atom thick.**

Membranes are fundamental components of a wide variety of physical, chemical, and biological systems, used in everything from cellular compartmentalization to mechanical pressure sensing. They divide space into two regions, each capable of possessing different physical or chemical properties. A simple example is the stretched surface of a balloon, where a pressure difference across the balloon is balanced by the surface tension in the membrane. Graphene, a single layer of graphite, is the ultimate limit: a chemically stable and electrically conducting membrane one atom in thickness[1-3]. An interesting question is whether such an atomic membrane can be impermeable to atoms, molecules and ions. In this letter, we address this question for gases. We show that these membranes are impermeable and can support pressure differences larger than one atmosphere. We use such pressure differences to tune the mechanical resonance frequency by ~100 MHz. This allows us to measure the mass and elastic constants of graphene membranes. We demonstrate that atomic layers of graphene have stiffness similar to bulk graphite ($E \sim$ 1 TPa). These results show that single atomic sheets can be integrated with microfabricated structures to create a new class of atomic scale membrane-based devices.

A schematic of the device geometry used here—a graphene-sealed microchamber—is shown in Fig. 1a. Graphene sheets are suspended over predefined wells in silicon oxide using mechanical exfoliation (see Supporting Information). Each graphene membrane is clamped on all sides by the van der Waals force between the graphene and SiO$_2$, creating a $\sim$ (µm)$^3$ volume of confined gas. The inset of Fig. 1a shows an optical image of a single layer graphene sheet forming a sealed square drumhead with a width $W$ = 4.75 µm on each side. Raman spectroscopy was used to confirm that this graphene sheet was a single layer in thickness[4-6]. Chambers with graphene thickness from 1 to $\sim$ 75 layers were studied.

After initial fabrication, the pressure inside the microchamber, $p_{int}$, is atmospheric pressure (101 kPa). If the pressure external to the chamber, $p_{ext}$, is changed, we found that $p_{int}$ will equilibrate to $p_{ext}$ on a time scale that ranges from minutes to days, depending on the gas species and the temperature. On shorter time scales than this equilibration time, a significant pressure difference $\Delta p = p_{int} - p_{ext}$ can exist across the membrane, causing it to stretch like the surface of a balloon (Fig. 1b). Examples are shown for $\Delta p > 0$ in Fig. 1c and $\Delta p < 0$ in Fig. 1d.

To create a positive pressure difference, $\Delta p > 0$, as shown in Fig. 1c, we place a sample in a pressure chamber with $p_{ext}$ = 690 kPa N$_2$ gas for 3 hours. After removing it, a tapping mode atomic force microscope (AFM) image at ambient external pressure (Fig. 1c) shows that the membrane bulges upwards. Similarly, we can create a lower pressure in the chamber, $\Delta p < 0$, by storing the device under vacuum and then returning it to atmospheric pressure. The graphene-sealed microchamber from Fig. 1a (inset) is placed in a pressure of $\sim$ 0.1 Pa for 4 days and then imaged in ambient conditions by AFM (Fig. 1d). The graphene membrane is now deflected downward indicating $p_{int} < p_{ext}$.

Over time, the internal and external pressures equilibrate. Figure 1e shows a series of AFM line traces through the center of the graphene membrane taken over a period of three days. The deflection $z$ at the center of the membrane is initially $z_o$ = 175 nm and decreases slowly over time, indicating a slow air leak from the microchamber. The time scale for decay is approximately 24 hours. We characterize the equilibration process by monitoring the pressure change and using the ideal gas law to convert this to a leak rate:

$$\frac{dN}{dt} = \frac{V}{k_B T} \frac{dp_{in}}{dt} \quad (1)$$

where $N$ is the number of atoms or molecules in the chamber. Figure 2 shows results for several different membranes of various thicknesses and for different gases. Air and argon show similar leak rates, while helium is 2 orders of magnitude faster. The helium leak rates ranged from $10^5$ atoms/s to $\sim 10^6$ atoms/s with no noticeable dependence on thickness from 1 – 75 atomic layers. All the data was taken in a similar manner where approximately the same pressure difference was applied across the membrane (see Supporting Information).

The lack of dependence of the leak rate on the membrane thickness indicates that the leak is not through the graphene sheets, or though defects in these sheets. This suggests it is either through the glass walls of the microchamber or through the graphene-$SiO_2$ sealed interface. The former can be estimated from the known properties of He diffusion through glass[7]. Using Fick's law of diffusion and typical dimensions for our microchambers we estimate a rate of ~ 1-5 × $10^6$ atoms/sec. This is close to the range of values measured (Fig. 2).

Using this measured leak rate, we estimate an upper bound for the average transmission probability of a He atom impinging on a graphene surface as:

$$\frac{dN}{dt} \frac{2d}{Nv} < 10^{-11} \quad (2)$$

where $d$ is the depth of the microchamber, and $v$ is the velocity of He atoms. In all likelihood, the true permeability is orders of magnitude lower than the bound given above. Simple estimates based on WKB tunneling of He atoms through a perfect graphene barrier (~ 8.7 eV barrier height, 0.3 nm thickness) and through a "window" mechanism whereby temporary bond breaking lowers the barrier height to ~ 3.5 eV, give a tunneling probability at room temperature many orders of magnitude smaller than we observe[8-10]. If we approximate Helium atoms as point particles, classical effusion through single atom lattice vacancies in the graphene membrane occurs in ~ 1 sec and therefore much faster than the rates we measure. We therefore conclude that the graphene layer is essentially perfect and for all intents and purposes impermeable to all standard gases, including He.

The impermeability of the graphene membrane allows us to use pressure differences to apply a large, well-defined force that is uniformly distributed across the entire surface of the membrane. This ability to create controlled strain in the membrane has many uses. First, we can measure the elastic properties of the graphene sheet. A well-known and reliable method used to study the elastic properties of films is the bulge test technique[11]. The deflection of a thin film is measured as a uniform pressure is applied across it. This surface tension, $S$, is the sum of two components: $S = S_0 + S_p$ where $S_0$ is the initial tension per unit length along the boundary and $S_p$ is the pressure-induced tension. Tension is directly related to the strain, $\varepsilon$, as $S = \frac{Et}{(1-\upsilon)}\varepsilon$, where $E$ is the Young's modulus, $t$ is the thickness, and $\upsilon$ is Poisson's ratio. For the geometry of a square membrane, the pressure difference as a function of deflection can be expressed as[11]:

$$\Delta p = \frac{4z}{W^2}\left(c_1 S_o + \frac{4c_2 Etz^2}{W^2(1-\upsilon)}\right) \qquad (3)$$

where $c_1 = 3.393$ and $c_2 = (0.8+0.062\upsilon)^{-3}$.

Using the deflection and pressure difference in Fig. 1d and accounting for initial slack in the membrane as discussed later in the text, we determine the elastic constants of graphene to be $Et/(1-\upsilon) = 390 \pm 20$ N/m (See Supporting Information). The accepted values for the experimental and theoretical elastic constants of bulk graphite and graphene—both 400 N/m[12-14]—are within the experimental error of our measurement. This is an important result in nanomechanics considering the vast literature examining the relevance of using elastic constants for bulk materials to describe atomic scale structures [12, 15].

The surface tension in the pressurized membrane can be readily obtained from the Young-Laplace equation, $\Delta p = S(1/R_x + 1/R_y)$ where $R_{x(y)}$ is the radius of curvature of the surface along the $x(y)$ direction. The shape of the bulged membrane with $\Delta p$ = -93 kPa in Fig. 1d directly gives $R_{x(y)}$. At the point of maximum deflection it is $R_x = R_y$ = 21 μm which amounts to a surface tension $S$ = 1 N/m. This is 14 times the surface tension of water, but corresponds to a small strain in the graphene of 0.26 %. The atomically thin sealed chambers reported here can support pressures up to a few atmospheres. Beyond this, we observe that the graphene slips on the surface. Improved clamping could increase allowable pressure differentials dramatically.

This pressure induced-strain in the membrane can also be used to control the resonance frequency of the suspended graphene. This is shown in Fig. 3a for a monolayer device prepared with a small gas pressure $p_{int}$ in the chamber. Figure 3b shows results on a 1.5 nm thick membrane. The vibrations of the membrane are actuated and measured optically, as previously reported[2]. The frequency changes dramatically with external pressure, exhibiting a sharp minimum at a specific pressure and growing on either side. Sufficiently far from the minimum frequency, $f_0$, the frequency scales as $f^3 \alpha \Delta p$ (Fig. 3b).

This behavior follows from the pressure induced changes in the tension $S$ in the membrane. Neglecting the bending rigidity, the fundamental frequency of a square membrane under uniform tension is given by:

$$f = \sqrt{\frac{S_0 + S_p}{2mW^2}} \qquad (4)$$

where $m$ is the mass per unit area[16]. Sufficiently far from $f_0$, equations (3) and (4) can be combined with the approximation, $S \approx \frac{\Delta p W^2}{16z}$ to get the following expression:

$$f^3 = \Delta p \sqrt{\frac{c_2 E t}{2048 m^3 W^4 (1-\upsilon)}} \qquad (5)$$

This gives the functional form observed in Fig. 3b with the prefactor consisting of the elastic constants of the membrane and the mass. Using $Et/(1-\upsilon)$ determined previously, we fit (5) to the data of Fig. 3a and 3b to determine the mass per area of the membranes. We find $m$ = 9.6 ± 0.6 x $10^{-7}$ kg/m² for the monolayer of Fig. 3a. This is 30 % higher than the theoretical value for a single layer of graphene of 7.4 x $10^{-7}$ kg/m$^2$. One possibility for this extra mass is adsorbates which would significantly shift the mass of a single atom membrane. The 1.5 nm-thick few-layer membrane of Fig. 3b has a $m$ = 3.1 ± 0.2 x $10^{-6}$ kg/m$^2$. This corresponds to ~ 4 atomic layers in thickness. Previous attempts to deduce the mass from resonance measurements of doubly clamped beams were obscured by the large initial tension in the resonators[2]. Exploiting the impermeability of graphene membranes to controllably tune the resonance frequency gives us the mass of the suspended graphene membrane regardless of this initial tension. To our knowledge, this is the first direct measurement of the mass of graphene.

The minimum frequency, $f_0$, corresponds to $S_p$ = 0, i.e. $p_{int}$ = $p_{ext}$. The monolayer graphene membrane in Fig. 3a has $f_0$ = 38 MHz when $\Delta p$ = 0. This frequency is significantly higher than expected for a graphene square plate under zero tension (0.3 MHz) suggesting that at $\Delta p$ = 0, the resonance frequency is dominated by $S_0$ and not the bending rigidity. Using the experimentally measured mass of the monolayer membrane above we deduce an $S_0$ ~ 0.06 N/m. This is similar to what was previously observed in doubly-clamped graphene beams fabricated by the same method[2].

The origin of this tension is clear from Fig. 4a which shows a tapping-mode AFM image of the suspended monolayer graphene membrane of Fig. 1d with $\Delta p$ = 0. The image shows the graphene membrane to have a ~ 17 nm dip along the edges of the suspended regions where the graphene meets the SiO$_2$ sidewalls (Fig. 4b). This results from the strong van der Waals interaction between the edge of the graphene membrane and the SiO$_2$ sidewalls (Fig. 4c), which previously has been estimated to be $U$ ~ 0.1 J/m$^2$ [17, 18]. This attraction yields a surface tension $S_0 = U$ ~ 0.1 N/m which is close to the value extracted from the resonance measurement.

The tension in the membrane can also be probed by pushing on the membrane with a calibrated AFM tip[19]. This force-deflection curve gives a direct measure of the spring constant $k_{graphene}$ = 0.2 N/m of the graphene membrane, as shown in Figure 4d. Neglecting the bending rigidity, the tension can be obtained using $S \approx (k_{graphene}/2\pi) \ln(R/r)$, where $R$ is the radius of the membrane and $r$ is the radius of the AFM tip[20]. Assuming $r$ ~ 50 nm, this gives $S$ ~ 0.1 N/m, close to both the theoretical value and the value measured using the resonance frequency technique above. These results show that self-tensioning in these thin graphene sheets dominates over the bending rigidity, and this tension will smooth corrugations that may occur in tension-free graphene membranes[3].

We envision many applications for these graphene sealed microchambers. They can act as compliant membrane sensors which probe pressures in small volumes and explore pressure changes associated with chemical reactions, phase transitions, and photon detection[21, 22]. In addition to these spectroscopic studies, graphene drumheads offer the opportunity to probe the permeability of gases through atomic vacancies in single layers of atoms[23] and defects patterned in the graphene membrane can act as selective barriers for ultrafiltration[24, 25]. The tensioned suspended graphene membranes also provide a platform for STM imaging of both graphene[26-28] and graphene-fluid interfaces and offer a unique separation barrier between 2 distinct phases of matter that is only one atom thick.

ACKNOWLEDGMENT: This work was supported by the NSF through the Cornell Center for Materials Research and Center for Nanoscale Systems and, and by the MARCO Focused Research Center on Materials, Structures, and Devices. Sample fabrication was performed at the Cornell node of the National Nanofabrication Users Network, funded by NSF. We thank Lihong Herman, Jiwoong Park, and Michael Jaquith for help with the spring constant calibration of the AFM tip. We also thank Alan Zehnder for useful discussion

**Supporting Information Available**: Experimental methods and a description of slack and self tensioning at $\Delta p = 0$.

FIGURE CAPTIONS:

Figure 1

(a) Schematic of a graphene sealed microchamber. (Inset) Optical image of a single atomic layer graphene drumhead on 440 nm of $SiO_2$. The dimensions of the microchamber are 4.75 μm x 4.75 μm x 380 nm. (b) Side view schematic of the graphene sealed microchamber. (c) Tapping mode atomic force microscope (AFM) image of a ~ 9 nm thick many layer graphene drumhead with $\Delta p > 0$. The dimensions of the square microchamber are 4.75 μm x 4.75 μm. The upward deflection at the center of the membrane is $z$ = 90 nm. (d) AFM image of the graphene sealed microchamber of Fig. 1a with $\Delta p$ = -93 kPa across it. The minimum dip in the z direction is 175 nm. (e) AFM line traces taken through the center of the graphene membrane of (a). The images were taken continuously over a span of 71.3 hours and in ambient conditions. (Inset) The deflection at the center of the graphene membrane vs. time. The first deflection measurement ($z$ = 175 nm) is taken 40 minutes after removing the microchamber from vacuum.

Figure 2

Scatter plot of the gas leak rates vs. thickness for all the devices measured. Helium rates are shown as solid triangles (▲), argon rates are shown as solid squares (■) and air rates are shown as hollow squares (□).

Figure 3

(a) Resonance frequency vs. external pressure for the single-layer graphene sealed microchamber shown in Fig. 1a. (Upper inset) Resonance frequency curve taken at $p_{ext}$ = 27 Pa with a resonance frequency of $f$ = 66 MHz and $Q$ = 25. (Lower insets) Schematic of the configuration of the microchamber at various applied pressures. The graphene is puffed upwards or downwards depending on $\Delta p$. (b) (upper) Resonance frequency vs. $p_{ext}$ for a 1.5 nm-thick few layer graphene sealed microchamber. Each curve was taken at a different time over a span of 207 hours, and the device was left in $p_{ext}$ ~ 0.1 mPa in between each measurement. (lower) (Resonance frequency)$^3$ vs. $p_{ext}$ for the red scan in Fig. 4b. A linear fit to the data is shown in red.

Figure 4

(a) Tapping mode AFM image of the single-layer graphene sealed microchamber shown in Fig. 1a with $\Delta p$ = 0. (b) Line cut through the center of the graphene membrane in (a). (c) Schematic of the graphene membrane at $\Delta p$ = 0 with an initial deflection $z_0$ due to self-tensioning. (d) Force-distance curve taken at the center of the graphene membrane in (a) at $\Delta p = 0$. The spring constant of the cantilever used is $k_{tip} =$ 0.67 N/m.

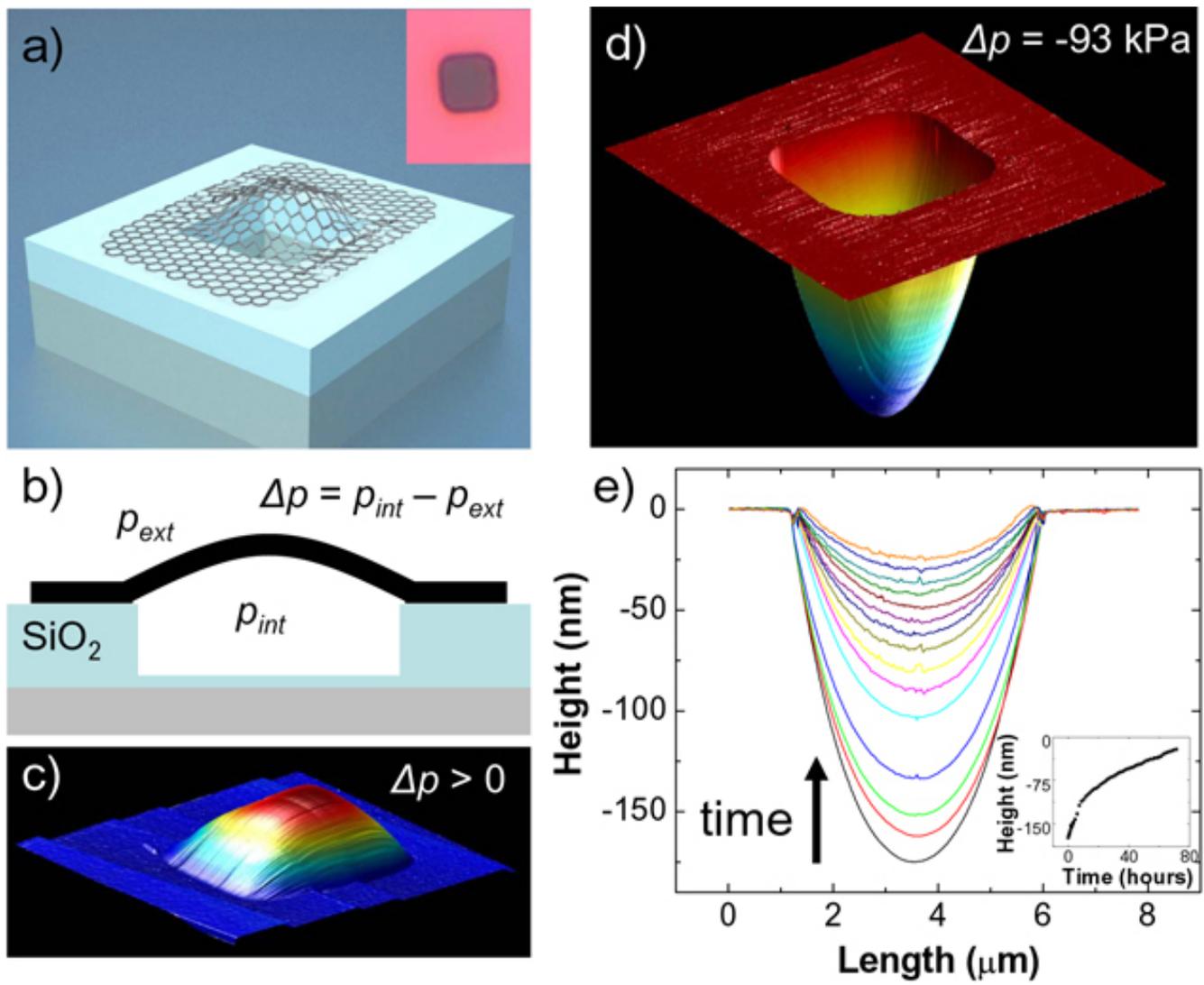

Figure 1

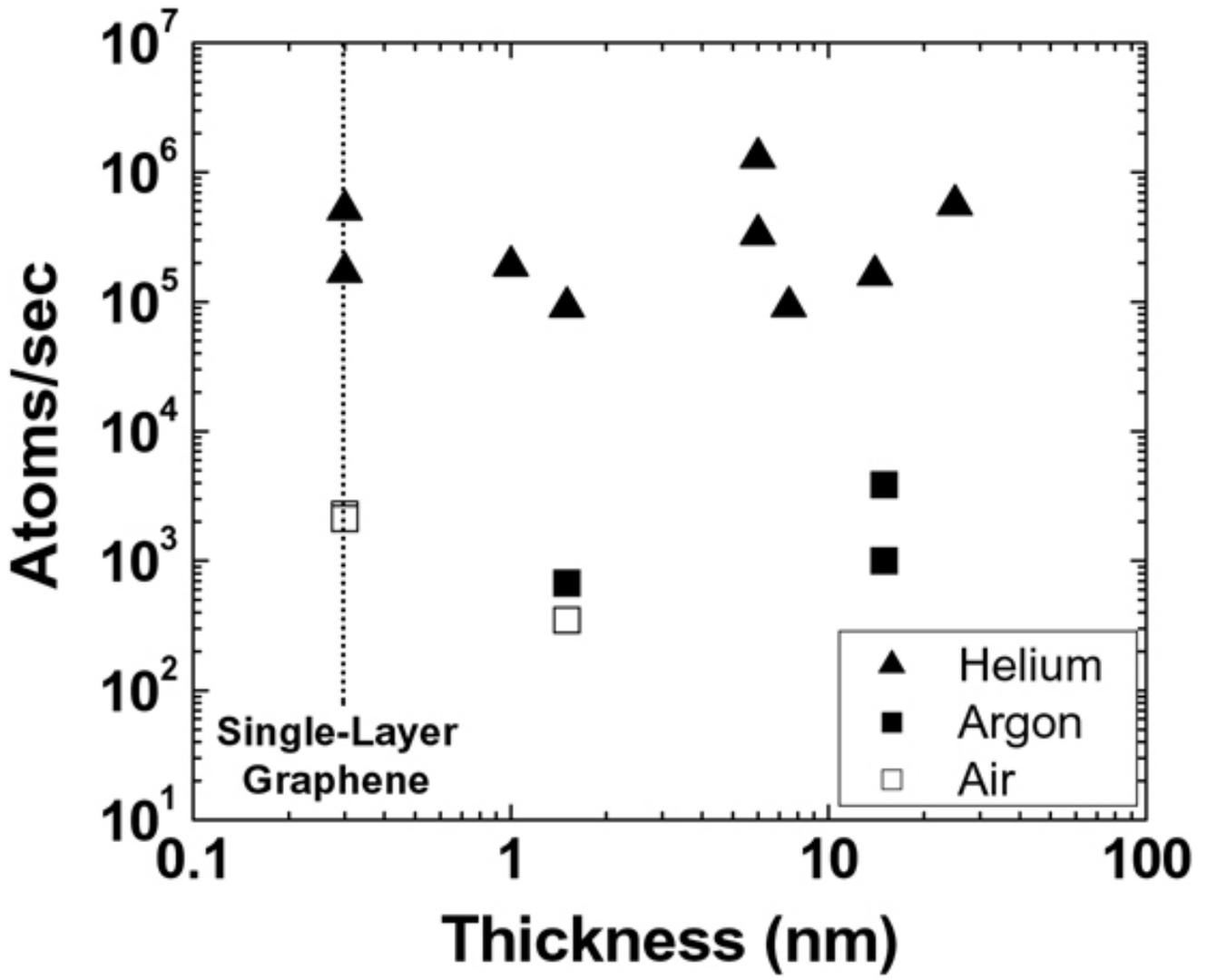

Figure 2

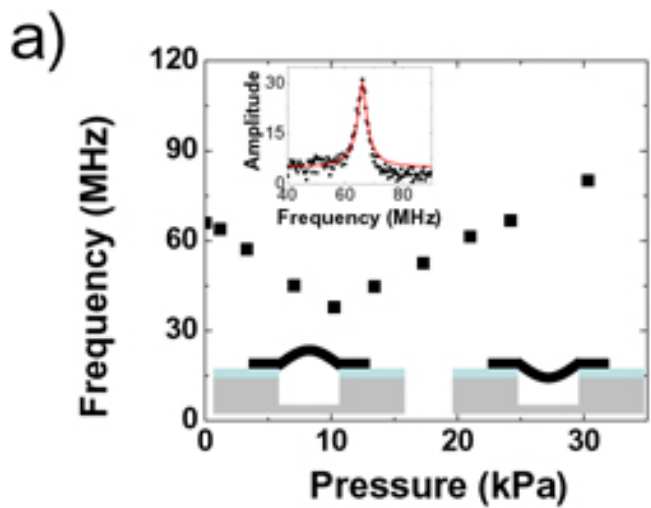
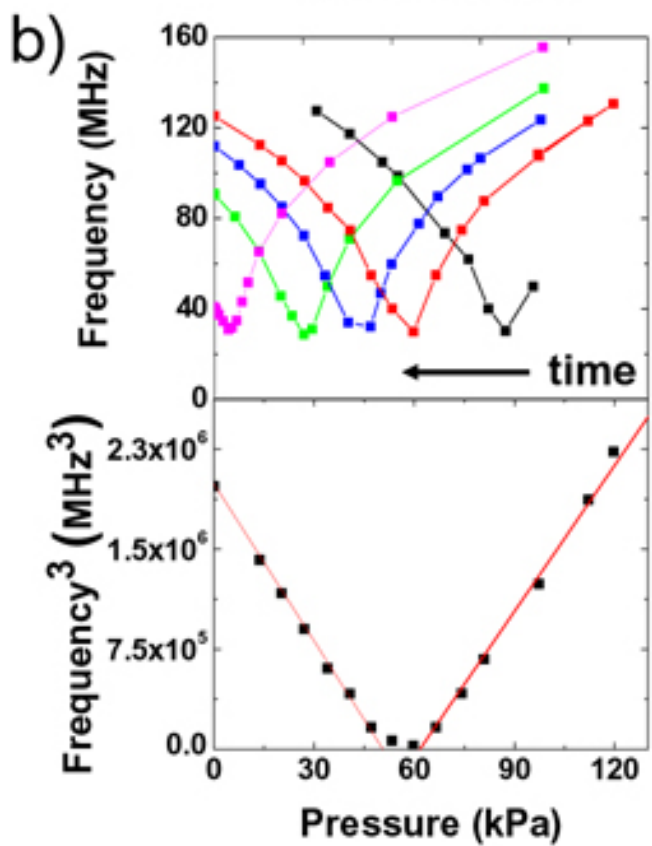

Figure 3

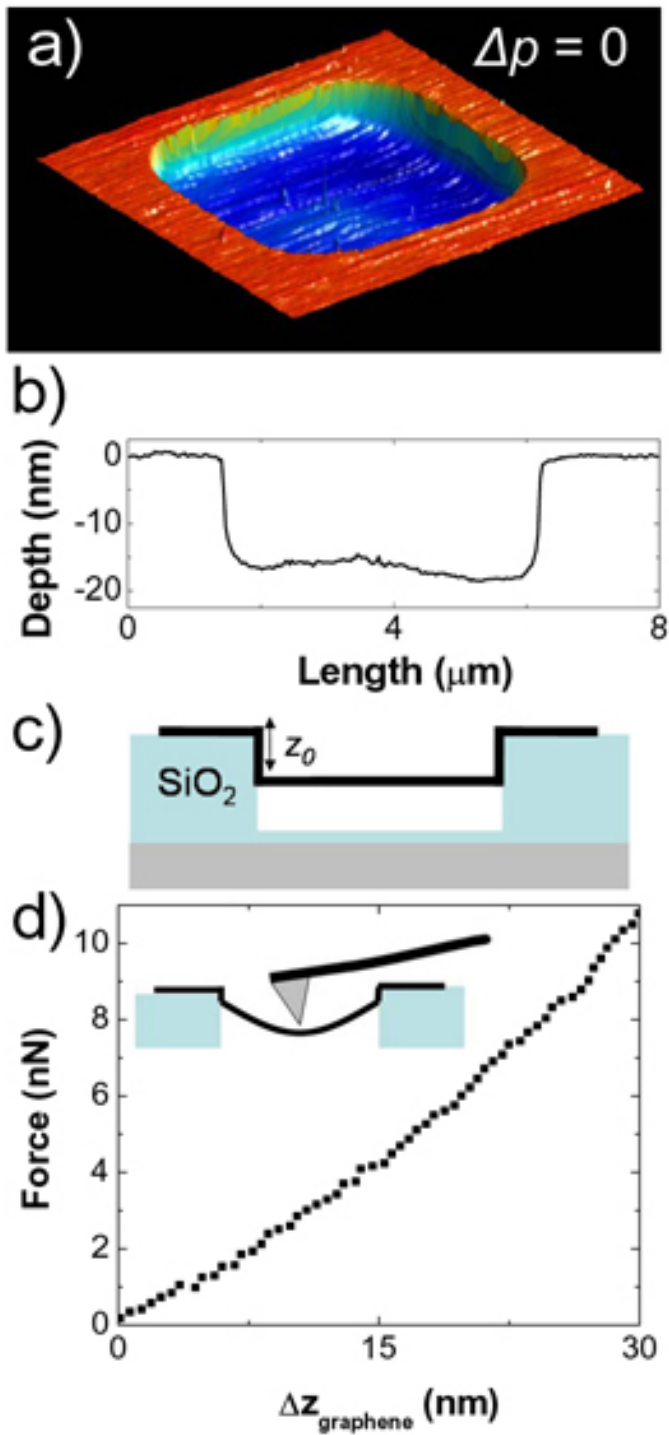

Figure 4

# Supplementary Information

## Impermeable Atomic Membranes from Graphene Sheets

J. Scott Bunch, Scott S. Verbridge, Jonathan S. Alden, Arend M. van der Zande, Jeevak M. Parpia, Harold G. Craighead, Paul L. McEuen

*Cornell Center for Materials Research, Cornell University, Ithaca NY 14853*

**Methods**

Graphene drumheads are fabricated by a combination of standard photolithography and mechanical exfoliation of graphene sheets. First, a series of squares with areas of 1 to 100 μm$^2$ are defined by photolithography on an oxidized silicon wafer with a silicon oxide thickness of 285 nm or 440 nm. Reactive ion etching is then used to etch the squares to a depth of 250 nm to 3 μm leaving a series of wells on the wafer. Mechanical exfoliation of Kish graphite using Scotch tape is then used to deposit suspended graphene sheets over the wells.

To determine the elastic constants of graphene using equation (3), we extrapolate the deflection in Fig. 1e (inset) to $z$ = 181 nm to account for a 40-minute sample-load time, assume an initial pressure difference across the membrane, $\Delta p$ = 93 kPa, and a negligible initial tension. The latter two assumptions are verified using resonance measurements. The actual deflection used in equation (3) is obtained by subtracting the extrapolated deflection $z$ = 181 nm from the initial deflection $z_0$ = 23 ± 3 nm at $\Delta p$ = 0. This initial deflection is determined from the AFM image in Fig. 4a and AFM force-distance curves Fig. S1.

The gas leak rate is measured by monitoring $p_{int}$ vs. time. For the case of the leak rate of air, the microchamber begins with $p_{int}$ ~ 100 kPa Air. This is verified by a scan of frequency vs. $p_{ext}$. A similar scan is performed once every few hours to monitor $p_{int}$ while the device is left at $p_{ext}$ ~ 0.1 mPa between each measurement (Fig. 3a and 3b). The leak rate of argon is measured in a similar manner except the microchamber begins with a $p_{int}$ ~ 0 kPa argon and ~ 10 kPa air. The microchamber is left in $p_{ext}$ ~ 100 kPa argon between measurements to allow argon to diffuse into the microchamber. This diffusion is monitored by finding the minimum pressure in a scan of frequency vs. $p_{ext}$.

To measure the helium leak rate we apply a $\Delta p \sim 40 – 50$ kPa He and monitor the resonance frequency as helium diffuses into the microchamber. It will diffuse until the partial pressure of helium is the same inside and outside the microchamber. From the slope of the line we extract a helium leak rate for the devices using equation (1). Leak rates from square membranes with sides varying from 2.5 to 4.8 µm were measured with no noticeable dependence of the leak rate on area.

**Slack and Self Tensioning at $\Delta p = 0$**

Since the cantilever-surface interaction is expected to be different for AFM measurements over the relatively-pliable suspended and the rigid SiO$_2$-supported graphene, the depth of the membrane $z_0$, at $\Delta p = 0$ must be determined via force and amplitude calibrations of the cantilever over each surface[1]. A representative calibration measurement is shown in Fig. S.1. Both the amplitude (upper) and deflection (lower) of the AFM tip is measured while approaching the surface.

Over the SiO$_2$-supported surface, the difference between the actual surface position and the position given by the image in Fig. 4a can be determined by subtracting the height at which the AFM tip begins to bend due to unbroken contact with the surface (A) from the height at which the amplitude setpoint intersects with the amplitude response curve (B) (Fig. S.1). The surface is determined to be 30 nm below the amplitude setpoint position.

Since suspended graphene is more pliable than supported graphene, the onset of the AFM cantilever's deflection of Fig. S.1 is more gradual, and thus cannot be readily used to determine the equilibrium height of the suspended graphene. Instead, we note that when in unbroken-contact with the graphene surface, any deviations of the AFM tip from the equilibrium (lowest-strain) depth of the membrane will result in an increase in the membrane tension as the tip either pulls up or pushes down on the membrane. This increase in tension on either side of the equilibrium position will cause a decrease in cantilever response amplitude, resulting in a peak in the cantilever-amplitude response at the equilibrium position, similar to what has been observed for suspended carbon nanotubes[1]. This occurs at ~100 nm, or 34 nm below the amplitude setpoint position (C).

Comparing these setpoint-to-surface depths for suspended and supported graphene, we find that the equilibrium depth of the suspended membrane is $17 + (34 – 30) = 21$ nm below the SiO$_2$-supported surface where 17 nm is the distance measured in Fig. 4a. Repeating these measurements across the center of the membrane yields an average equilibrium membrane-depth depth $z_0 = 17 \pm 1$ nm + ($6 \pm 2$ nm) = $23 \pm 3$ nm.

Supplementary References:

Figure S1

(upper) Driven oscillation amplitude of the tapping mode AFM cantilever with resonance frequency = 349 kHz vs. piezo extension as tip is brought into contact with the surface. Black and red are extension and retraction curves over the supported graphene on SiO2 surface. Green and blue are extension and retraction curves over the suspended graphene membrane. (lower) The deflection of the cantilever vs. piezo extension. The upper and lower traces were taken simultaneously.

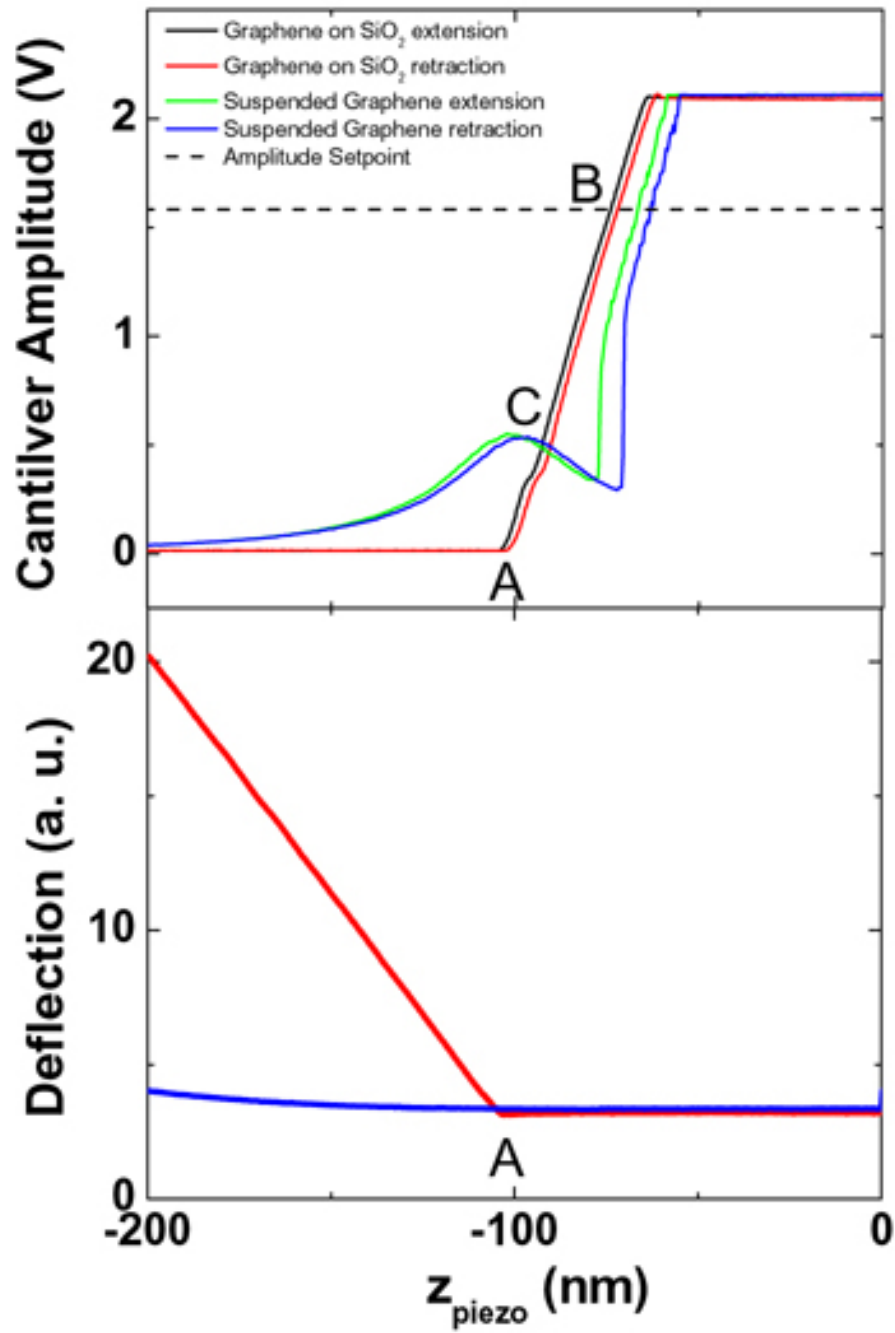

Figure S1